\documentclass[aps,prl,showpacs,floatfix,twocolumn]{revtex4}

\usepackage{amsmath,bm}
\usepackage[dvipdfm]{graphicx}

\usepackage[colorlinks=true,linkcolor=blue,plainpages=false]{hyperref}

\def\v2{\mbox{$v_2$}}

\begin{document}

%
\hyphenation{author another created financial paper re-commend-ed Post-Script}

\title{ An estimate for the location of QCD critical end point }
\author{ Roy~A.~Lacey} 
\author{ N.~N.~Ajitanand} 
\author{ J.~M.~Alexander}
\author{ P.~Chung}
\author{ J.~Jia}
\author{A.~Taranenko}
\affiliation{ Department of Chemistry, State University of New York at Stony Brook, Stony 
Brook, NY 11794-3400, USA
} 
\author{P.~Danielewicz}
\affiliation{National Superconducting Cyclotron Laboratory and
Department of Physics and Astronomy, Michigan State University, 
East Lansing, MI 48824-1321, USA.
}
\date{\today}

\begin{abstract}
	It is proposed that a study of the ratio of shear viscosity to entropy 
density $\frac{\eta}{s}$ as a function of the baryon chemical potential $\mu_B$, 
and temperature T, provides a dynamic probe for the critical end point (CEP) in 
hot and dense QCD matter. An initial estimate from an elliptic flow excitation 
function gives $\mu^{\text{cep}}_B \sim 150-180$~MeV and $T_{\text{cep}} \sim 165 - 170$~MeV 
for the location of the the CEP. These values place the CEP in the range for ``immediate" 
validation at RHIC. 
\end{abstract}

\maketitle


	The phase boundaries and the critical end point (CEP) are fundamental 
characteristics of hot and dense nuclear matter \cite{Asakawa:1989bq}. 
The study of heavy ion collisions has been proposed  \cite{Stephanov:1998dy} as 
an avenue to search for these essential characteristics of the Quantum 
Chromodynamics (QCD) phase diagram i.e. the plane of temperature vs baryon chemical 
potential ($T, \mu_B$).

	A recent resurgence of experimental interest in the CEP has been aided by  
strong experimental and theoretical evidence for a crossover transition to 
the quark gluon plasma (QGP) in heavy ion collisions at the Relativistic 
Heavy Ion collider (RHIC) \cite{Adcox:2004mh,Adams:2005dq,Back:2004je,Arsene:2004fa,
Fodor:2001pe,Gyulassy:2004zy,Muller:2004kk,Shuryak:2004cy,Heinz:2001xi}.
Such a crossover, constitutes a necessary requirement, albeit insufficient, 
for locating the CEP.

	Several attempts have been made to provide theoretical guidance on where to 
localize a search for the CEP in the QCD phase 
diagram \cite{Fodor:2001pe,deForcrand:2003hx,Allton:2005gk,Gavai:2004sd,
Philipsen:2005mj}. The resulting predictions for the critical values 
of temperature $T_{\text{cep}}$ and baryon chemical potential $\mu^{\text{cep}}_B$, 
which locates the CEP, have not converged and now span a broad range. 

\begin{figure}[tb]
 \includegraphics[width=1.0\linewidth]{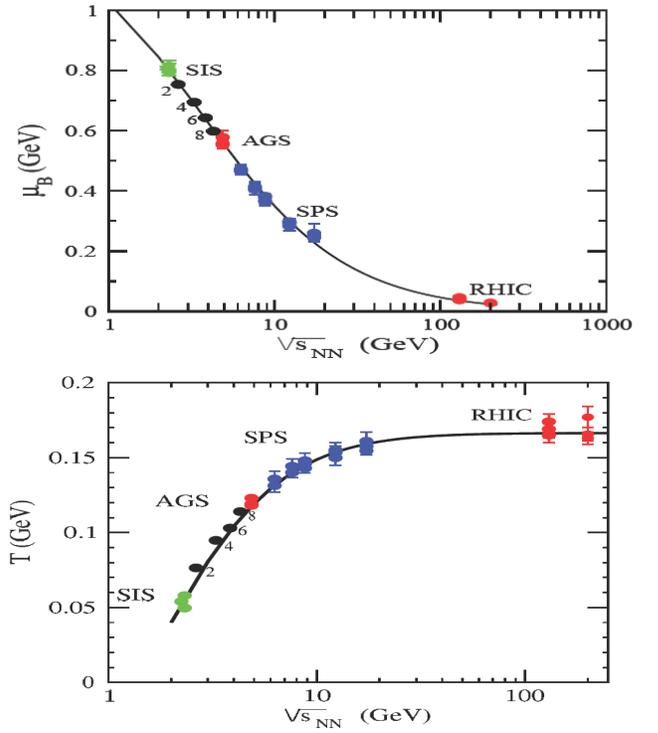}  
 \caption{\label{fig1}
	(Color online) $\sqrt{s_{NN}}$ dependence of $\mu_B$ (top panel) and $T$ obtained 
	from chemical fits \cite{Cleymans:2005xv} to particle yield ratios obtained at different 
	accelerator facilities as indicated. The solid lines are fits to these data.
}
\end{figure}

	Therefore, recent plans for the experimental verification 
of the CEP have centered on energy scans with an eye toward accessing 
the broadest possible range of $\mu_B$ and $T$ values \cite{CpWorkshop:2006,Stephans:2006tg,
Gazdzicki:1998vd,Gazdzicki:2005gs}. Fig. \ref{fig1} reenforces the value of such 
energy scans; it shows the chemical freezeout values of $\mu_B$ (top panel) and 
$T$, as a function of beam collision energy $\sqrt{s_{NN}}$, extracted 
via chemical fits to particle ratios \cite{Cleymans:2005xv} obtained at several accelerator 
facilities. This unprecedented reach in $\mu_B$ and $T$ values, clearly indicate that 
the combined results from energy scans at the Facility for Anti-proton and Ion Research (FAIR),
the Super Proton Synchrotron (SPS) and RHIC, will allow access to the full range of $\mu_B$ and $T$ 
values necessary for a comprehensive CEP search.
%
%

	At the CEP (or close to it) anomalies can occur in a wide variety of dynamic 
and static properties. Anomalies in dynamic properties reflect a change in quantities 
such as the transport coefficients and relaxation rates, multi-time correlation functions and  
the linear response to time-dependent perturbations. All of these depend on the 
equations of motion, and are not simply determined by the equilibrium distribution of 
the particles at a given instant of time. By contrast, static properties are 
solely determined by the single-time equilibrium distribution which include thermodynamic 
coefficients, single-time correlation functions, and the linear response to time-independent
perturbations.

	Critical fluctuations are thought to be one of the more important  
static signals for locating the CEP \cite{Costa:2007nc}. Consequently, extensive 
studies of particle fluctuations have been made over a broad range of beam collision 
energies. To date, no definitive observation indicative of the CEP, has been reported. 
A comprehensive search for the CEP via dynamic variables is still lacking. 

	In this letter, we argue that it is possible to locate the CEP via study of the $\mu_{B}$ and $T$ 
dependence of the ratio of viscosity to entropy density 
($\frac{\eta}{s}$) \cite{Csernai:2006zz,Lacey:2006bc,Chen:2006ig}    
and give an estimate for $\mu^{\text{cep}}_B$ and $T_{\text{cep}}$ (i.e the location of the CEP) 
using existing data.
	
\begin{figure}[tb]
 \includegraphics[width=1.0\linewidth]{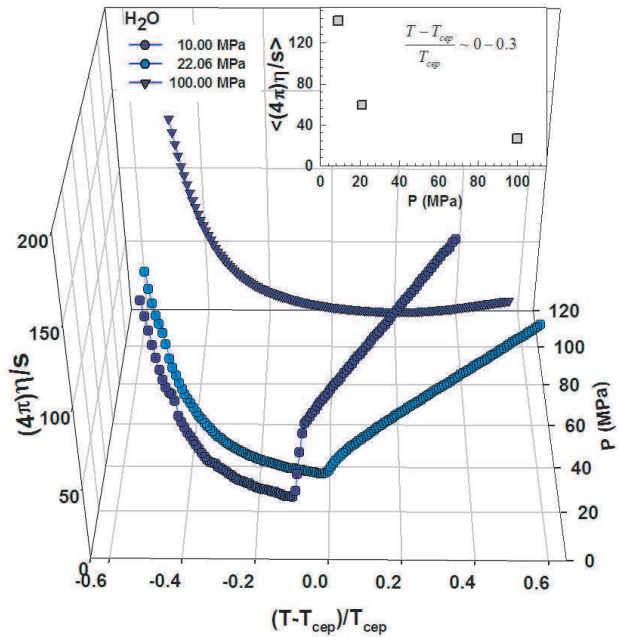}  
 \caption{\label{fig2}
	(Color online) $4\pi(\frac{\eta}{s})$ vs the reduced 
	temperature $\frac{T-T_{\text{cep}}}{T_{\text{cep}}}$ for H$_2$O. Results are shown for an 
	isobar at the critical pressure ($P_{\text{cep}} = 22.6$ MPa) and one above (below) it, 
	as indicated. The inset shows $\left\langle 4\pi(\frac{\eta}{s})\right\rangle$ for the 
	reduced temperature range $0.0-0.3$. The data are taken from 
	Ref.~\cite{Csernai:2006zz}.
}
\end{figure}

	The rationale for using $\frac{\eta}{s}$ as a probe for the CEP is 
two fold. First, we observe that this QCD critical endpoint belongs to the universality 
class of the 3d Ising model, ie. the same universality class for a liquid-gas system;
here, it is important to recall that all members of a given universality 
class have ``identical'' critical properties.
Second, we observe that $\frac{\eta}{s}$, for atomic and molecular substances, 
exhibits a minimum of comparable depth for different isobars passing in the vicinity 
of the liquid-gas critical end point \cite{Kovtun:2004de,Csernai:2006zz,Lacey:2006bc,Chen:2006ig}.
Fig.~\ref{fig2} illustrates this for H$_2$O. 
When an isobar passes through the critical end point, 
the minimum forms a cusp at the reduced temperature $\frac{T-T_{\text{cep}}}{T_{\text{cep}}}=0$; 
when it passes above the critical end point (i.e a pressure $P$ above the 
critical pressure $P_{\text{cep}}$), a less pronounced 
minimum is found at a value slightly above $\frac{T-T_{\text{cep}}}{T_{\text{cep}}}=0$.
For an isobar passing just below the critical pressure, the minimum is found at 
$\frac{T-T_{\text{cep}}}{T_{\text{cep}}} < 0$ (liquid side) but is accompanied by a 
discontinuous change across the phase transition. Thus, for a range of reduced 
temperatures, the average value $\left\langle 4\pi(\frac{\eta}{s})\right\rangle$ 
can be seen to grow rapidly for isobars passing through the critical end point
and just below it. This is illustrated in the inset of 
Fig.~\ref{fig2} where $\left\langle 4\pi(\frac{\eta}{s})\right\rangle$ 
is plotted vs pressure for the reduced temperature range 
$\frac{T-T_{\text{cep}}}{T_{\text{cep}}}=0-0.3$. Fig.~\ref{fig2} shows that the 
CEP is signaled by a minimum at $\frac{T-T_{\text{cep}}}{T_{\text{cep}}}\sim 0$, in the 
dependence of $4\pi(\frac{\eta}{s})$ on the reduced temperature, as well as   
a sharp increase in $\left\langle 4\pi(\frac{\eta}{s})\right\rangle$ vs P 
for $\frac{T-T_{\text{cep}}}{T_{\text{cep}}} \agt 0$.

		In analogy to the observations for atomic and molecular substances, one  
expects a range of trajectories, in the $(T,\mu_B)$ plane 
for decaying nuclear systems, to show $\frac{\eta}{s}$ minima with a possible 
cusp at the critical end point $(T_{\text{cep}},\mu^{{\text{cep}}}_B)$. 
That is, for $ \mu_B = \mu^{\text{cep}}_B$ the $\frac{\eta}{s}$ minimum is 
expected at the reduced temperature $\frac{T-T_{\text{cep}}}{T_{\text{cep}}} = 0$; 
for other values of $\mu_B$ with associated critical temperature $T_c$ not too 
far from $T_{\text{cep}}$, the dependence of $4\pi(\frac{\eta}{s})$ 
on $\frac{T-T_{\text{cep}}}{T_{\text{cep}}}$ 
is also expected to grow stronger as $\mu_B$ is increased from an 
initially small value up to $\mu_B \agt \mu^{\text{cep}}_B$.
Indeed, recent calculations for different types of phase transitions
(first-order, second-order and a crossover) suggest a rapid change 
in the value of $\eta/s$ in the vicinity of the CEP \cite{Chen:2007jq}.


	For a given value of $\mu_B$, a hot nuclear system for 
which $\frac{T-T_{\text{cep}}}{T_{\text{cep}}} > 0$
will sample the full range of $\frac{\eta}{s}$ values to give an average, as it  
evolves toward the $\frac{\eta}{s}$ minimum. Consequently, 
one expects an increase of $\left\langle 4\pi(\frac{\eta}{s})\right\rangle$ with 
increasing $\mu_B$, punctuated by a relatively rapid increase 
for $\mu_B$ values slightly above $\mu^{\text{cep}}_B$. The latter would be comparable to 
the rapid increase in $\left\langle 4\pi(\frac{\eta}{s})\right\rangle$
observed for H$_2$O (inset in Fig.~\ref{fig2}) when $P$ is lowered a little  
below the critical pressure. 

	Therefore, the extraction of $\left\langle 4\pi(\frac{\eta}{s}) \right\rangle$ as 
a function of $T$ and $\mu_B$ from experimental data, can serve as a constraint for the 
location of the CEP. Such extractions are possible from an elliptic flow excitation 
function measurement because a sizable change in $\left\langle 4\pi(\frac{\eta}{s}) \right\rangle$ 
is expected to lead to a measurable suppression of the magnitude of 
elliptic flow. It could even serve to invalidate the currently 
observed universal scaling patterns \cite{Issah:2006qn,Adare:2006ti,Lacey:2006pn}.

	In recent work \cite{Lacey:2006bc}, we have used elliptic flow 
measurements to obtain the estimates 
$\left\langle 4\pi(\frac{\eta}{s}) \right\rangle  \sim 1.3$ \cite{etas_uncertainty}
and $\left\langle T \right\rangle \sim 165$~MeV  for hot and dense 
matter \cite{compare_etas} produced in Au+Au collisions 
($\sqrt{s_{NN}}=200$~GeV \text{or} $\mu_B \sim 24$~MeV) at RHIC. 
A comparison of this $\frac{\eta}{s}$ value to those calculated for a meson-gas for 
$ T < T^Q_c$ \cite{Chen:2006ig}, and the QGP for $T > T^Q_c$ 
($T^Q_c \sim 170$ MeV \cite{Karsch:2000kv}), gave a good 
indication for the expected minimum (for $T$ close to $T_{\text{cep}}$) in the 
plot of $4\pi(\frac{\eta}{s})$ vs $\frac{T-T^Q_c}{T^Q_c}$. 
We therefore use this observation as a basis for the estimate 
$T_{\text{cep}} \sim 165 - 170$~MeV. This estimate is similar to the chemical 
freeze-out temperature for a broad range of collision energies (see botom panel 
of Fig. \ref{fig1}). This constancy of the freeze-out temperature ($T \sim 165$~MeV) may be 
a further indication that chemical freeze-out occurs at, or close to 
$T_{\text{cep}}$ for $\sqrt{s_{NN}} \sim 17-200$~GeV.

 The value $\left\langle 4\pi(\frac{\eta}{s}) \right\rangle  \sim 1.3$, 
achieved in Au+Au collisions at $\sqrt{s_{NN}}=200$~GeV, is rather close 
to the conjectured lower bound of $4\pi(\frac{\eta}{s})=1.0$. 
Consequently, one can conclude that, for $\mu_B \sim 24$~MeV, 
the hot expanding system spends a considerable portion of its dynamics in the region of 
low $\frac{\eta}{s}$; an optimal situation being, the system stays at low $\frac{\eta}{s}$ 
and then very quickly freezes out at or close to $T_{\text{cep}}$. Such a trajectory would be 
tantamount to the isobaric trajectory above the critical pressure, 
shown for H$_2$O in Fig.~\ref{fig2} (triangles). 
For other trajectories with $\mu_B$ close to, or slightly above $\mu^{\text{cep}}_B$, 
significant collective motion is expected to develop as the 
system evolves toward freeze-out with higher $\frac{\eta}{s}$ values. 
Consequently, the dependence of elliptic flow on $\mu_B$ is of interest.
%

\begin{figure}[t]
 \includegraphics[width=1.0\linewidth]{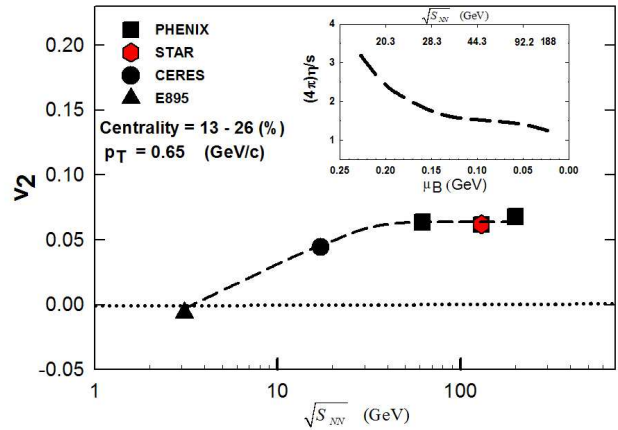}  
 \caption{\label{fig3}
	(Color online) Flow excitation function. The data is obtained from Ref. \cite{Adler:2004cj}.
	The inset shows the schematic variation of $\left\langle 4\pi(\frac{\eta}{s})\right\rangle$ vs $\mu_B$.
}
 \end{figure}

	Figure \ref{fig3} shows a differential elliptic flow ($v_2$) excitation function for 
charged hadrons, $\left\langle p_T \right\rangle = 0.65$~GeV/c, measured in 
$13-26$\% central Au+Au and Pb+Pb collisions \cite{Adler:2004cj}. 
For the range of collision energies 
$\sqrt{s_{NN}} \sim 17-200$~GeV, Fig. \ref{fig1} indicates an essentially constant 
freeze-out temperature $T \sim 165$~MeV for the values $\mu_B \sim 25 - 250$~MeV.
Therefore, these $v_2$ measurements (for $\sqrt{s_{NN}} \sim 17-200$~GeV) result from 
excited systems which all evolve toward our assumed value of $T_{\text{cep}}$, albeit 
with different $\mu_B$ values.  

	Figure \ref{fig3} shows that $v_2$ is essentially constant for 
$\sqrt{s_{NN}} \sim 62-200$~GeV. The energy density is estimated to decrease by 
$\sim 30$\% as the beam collison energy is reduced from $\sqrt{s_{NN}} \sim 200$~GeV 
to $\sqrt{s_{NN}} \sim 62$~GeV. Therefore we interpret this constancy of $v_2$ 
as an indication that (i) the equation of state associated with the crossover transition 
to the QGP is soft, and (ii) that $\left\langle \frac{\eta}{s} \right\rangle$ is relatively 
small for the $\mu_B$ values corresponding to this collision energy range.
That is, these $\mu_B$ values are significantly smaller than $\mu^{\text{cep}}_B$.

	For $\sqrt{s_{NN}} \sim 18$~GeV Fig. \ref{fig3} shows that $v_2$ decreases 
by almost 50\%, compared to the value for $\sqrt{s_{NN}} \sim 62-200$~GeV.
Here, it is important to point out that the mean transverse energy per particle is 
essentially the same for collision energy range $\sqrt{s_{NN}} \sim 17-200$~GeV and 
the estimated Bjorken energy density is $\sim 5.4$ and $3.2$ GeV/$\text{fm}^3$ for 
$\sqrt{s_{NN}} = 200$~GeV \cite{Adler:2004zn} and $\sqrt{s_{NN}} = 17$~GeV \cite{Margetis:1994tt} 
respectively, i.e the energy density change from $\sqrt{s_{NN}} \sim 62$~GeV 
to $\sqrt{s_{NN}} \sim 17$~GeV is not very large. Thus, the initial temperature 
of the high energy density matter created in collisions at $\sqrt{s_{NN}} \sim 62$~GeV 
and $\sqrt{s_{NN}} \sim 17$~GeV are not drastically different, and the 
fraction of the elliptic flow generated during the (dissipative) hadronic 
phase \cite{Hirano:2005xf} is expected to be qualitatively similar.

	A reduction in collision energy from $\sqrt{s_{NN}} \sim 62$~GeV 
to $\sqrt{s_{NN}} \sim 17$~GeV leads to a significant increase (more than a factor of two) 
in the value of $\mu_B$. Recent calculations \cite{Itakura:2007mx} also indicate that,
in the hadronic phase, $\frac{\eta}{s}$ decreases with increasing $\mu_B$. 
Therefore, a significant part of the reduction in $v_2$ observed as the collision energy 
is reduced from  $\sqrt{s_{NN}} \sim 62$~GeV to $\sqrt{s_{NN}} \sim 18$~GeV, 
could be a manifestation of the expected increase in 
$\left\langle \frac{\eta}{s} \right\rangle$ for values of $\mu_B \agt \mu^{\text{cep}}_B$.

	To estimate $\mu^{\text{cep}}_B$, we assume a smooth transition in the magnitude 
of $v_2$ over the range $62 \alt \sqrt{s_{NN}} \agt 18$~GeV, which is not yet measured. The 
inset in Fig. \ref{fig3} gives a schematic illustration of the expected change 
of $\left\langle 4\pi(\frac{\eta}{s}) \right\rangle$ with $\mu_B$ over this 
collision energy range. We use the ``knee" in the extrapolated values for 
$v_2$ over the range $62 \alt \sqrt{s_{NN}} \agt 18$~GeV to obtain the estimate 
$\mu^{\text{cep}}_B \sim 150-180$~MeV. A similar estimate was obtained by evaluating the 
$\mu_B$ dependence of $\left\langle \frac{\eta}{s} \right\rangle$ 
following the procedures outlined in Refs. \cite{Lacey:2006bc,Csernai:2006zz,Landdau1986,Eckart1940}, 
followed by interpolation to unmeasured $\mu_B$ values.

%

	Given these values of $T_{\text{cep}}$ and $\mu^{\text{cep}}_B$, we expect the 
value $\left\langle 4\pi(\frac{\eta}{s}) \right\rangle$, extracted from 
flow measurements performed at $\sqrt{s_{NN}} \sim 40$ and $30$~GeV, to 
be significantly larger than those obtained from similar measurements 
performed over the range $\sqrt{s_{NN}} \sim 62-200$~GeV. This change 
should also be reflected in the onset of a decrease of $v_2$, a possible 
violation of the universal scaling 
patterns \cite{Issah:2006qn,Adare:2006ti,Lacey:2006pn} observed 
for measurements at $\sqrt{s_{NN}} \sim 62-200$~GeV and a measurable 
increase in $v_2$ fluctuations.

	In summary we have argued that experimental assessment of 
$\left\langle \frac{\eta}{s} \right\rangle$ as a function of $\mu_B$ and $T$ 
provides a good dynamic observable for constraining the critical end point of  
hot QCD matter. A first estimate for the CEP from flow data indicate the 
values $T_{\text{cep}} \sim 165-170$ and $\mu^{\text{cep}}_B \sim 150-180$~MeV. 
Interestingly, our estimate is in good agreement with the prediction of 
Gavai et al. \cite{Gavai:2004sd}, obtained from lattice QCD 
simulations with realistic (about 1.7 times) pion masses and large volumes.
This estimate also places the CEP in the range for direct validation at RHIC via an  
energy scan. An initial measurement at  $\sqrt{s_{NN}} \sim 40$ and $30$~GeV 
would give sufficient information on where to focus more detailed attention.  

This work was supported by the US DOE under contract
DE-FG02-87ER40331.A008 and by the U.S. National Science 
Foundation under Grant No. PHY-0555893.
%
%
\bibliography{Critical_Point}

\begin{thebibliography}{45}
\expandafter\ifx\csname natexlab\endcsname\relax\def\natexlab#1{#1}\fi
\expandafter\ifx\csname bibnamefont\endcsname\relax
  \def\bibnamefont#1{#1}\fi
\expandafter\ifx\csname bibfnamefont\endcsname\relax
  \def\bibfnamefont#1{#1}\fi
\expandafter\ifx\csname citenamefont\endcsname\relax
  \def\citenamefont#1{#1}\fi
\expandafter\ifx\csname url\endcsname\relax
  \def\url#1{\texttt{#1}}\fi
\expandafter\ifx\csname urlprefix\endcsname\relax\def\urlprefix{URL }\fi
\providecommand{\bibinfo}[2]{#2}
\providecommand{\eprint}[2][]{\url{#2}}

\bibitem[{\citenamefont{Asakawa and Yazaki}(1989)}]{Asakawa:1989bq}
\bibinfo{author}{\bibfnamefont{M.}~\bibnamefont{Asakawa}} \bibnamefont{and}
  \bibinfo{author}{\bibfnamefont{K.}~\bibnamefont{Yazaki}},
  \bibinfo{journal}{Nucl. Phys.} \textbf{\bibinfo{volume}{A504}},
  \bibinfo{pages}{668} (\bibinfo{year}{1989}).

\bibitem[{\citenamefont{Stephanov et~al.}(1998)\citenamefont{Stephanov,
  Rajagopal, and Shuryak}}]{Stephanov:1998dy}
\bibinfo{author}{\bibfnamefont{M.~A.} \bibnamefont{Stephanov}},
  \bibinfo{author}{\bibfnamefont{K.}~\bibnamefont{Rajagopal}},
  \bibnamefont{and} \bibinfo{author}{\bibfnamefont{E.~V.}
  \bibnamefont{Shuryak}}, \bibinfo{journal}{Phys. Rev. Lett.}
  \textbf{\bibinfo{volume}{81}}, \bibinfo{pages}{4816} (\bibinfo{year}{1998}),
  \eprint{hep-ph/9806219}.

\bibitem[{\citenamefont{Adcox et~al.}(2005)}]{Adcox:2004mh}
\bibinfo{author}{\bibfnamefont{K.}~\bibnamefont{Adcox}} \bibnamefont{et~al.},
  \bibinfo{journal}{Nucl. Phys.} \textbf{\bibinfo{volume}{A757}},
  \bibinfo{pages}{184} (\bibinfo{year}{2005}), \eprint{nucl-ex/0410003}.

\bibitem[{\citenamefont{Adams et~al.}(2005)}]{Adams:2005dq}
\bibinfo{author}{\bibfnamefont{J.}~\bibnamefont{Adams}} \bibnamefont{et~al.},
  \bibinfo{journal}{Nucl. Phys.} \textbf{\bibinfo{volume}{A757}},
  \bibinfo{pages}{102} (\bibinfo{year}{2005}), \eprint{nucl-ex/0501009}.

\bibitem[{\citenamefont{Back et~al.}(2005)}]{Back:2004je}
\bibinfo{author}{\bibfnamefont{B.~B.} \bibnamefont{Back}} \bibnamefont{et~al.},
  \bibinfo{journal}{Nucl. Phys.} \textbf{\bibinfo{volume}{A757}},
  \bibinfo{pages}{28} (\bibinfo{year}{2005}), \eprint{nucl-ex/0410022}.

\bibitem[{\citenamefont{Arsene et~al.}(2005)}]{Arsene:2004fa}
\bibinfo{author}{\bibfnamefont{I.}~\bibnamefont{Arsene}} \bibnamefont{et~al.},
  \bibinfo{journal}{Nucl. Phys.} \textbf{\bibinfo{volume}{A757}},
  \bibinfo{pages}{1} (\bibinfo{year}{2005}), \eprint{nucl-ex/0410020}.

\bibitem[{\citenamefont{Fodor and Katz}(2002)}]{Fodor:2001pe}
\bibinfo{author}{\bibfnamefont{Z.}~\bibnamefont{Fodor}} \bibnamefont{and}
  \bibinfo{author}{\bibfnamefont{S.~D.} \bibnamefont{Katz}},
  \bibinfo{journal}{JHEP} \textbf{\bibinfo{volume}{03}}, \bibinfo{pages}{014}
  (\bibinfo{year}{2002}), \eprint{hep-lat/0106002}.

\bibitem[{\citenamefont{Gyulassy et~al.}(2005)}]{Gyulassy:2004zy}
\bibinfo{author}{\bibfnamefont{M.}~\bibnamefont{Gyulassy}}
  \bibnamefont{et~al.}, \bibinfo{journal}{Nucl. Phys.}
  \textbf{\bibinfo{volume}{A750}}, \bibinfo{pages}{30} (\bibinfo{year}{2005}),
  \eprint{nucl-th/0405013}.

\bibitem[{\citenamefont{M{\"{u}}ller}(2004)}]{Muller:2004kk}
\bibinfo{author}{\bibfnamefont{B.}~\bibnamefont{M{\"{u}}ller}}
  (\bibinfo{year}{2004}), \eprint{nucl-th/0404015}.

\bibitem[{\citenamefont{Shuryak}(2005)}]{Shuryak:2004cy}
\bibinfo{author}{\bibfnamefont{E.~V.} \bibnamefont{Shuryak}},
  \bibinfo{journal}{Nucl. Phys.} \textbf{\bibinfo{volume}{A750}},
  \bibinfo{pages}{64} (\bibinfo{year}{2005}), \eprint{hep-ph/0405066}.

\bibitem[{\citenamefont{Heinz et~al.}(2002)}]{Heinz:2001xi}
\bibinfo{author}{\bibfnamefont{U.}~\bibnamefont{Heinz}} \bibnamefont{et~al.},
  \bibinfo{journal}{Nucl. Phys.} \textbf{\bibinfo{volume}{A702}},
  \bibinfo{pages}{269} (\bibinfo{year}{2002}).

\bibitem[{\citenamefont{de~Forcrand and Philipsen}(2003)}]{deForcrand:2003hx}
\bibinfo{author}{\bibfnamefont{P.}~\bibnamefont{de~Forcrand}} \bibnamefont{and}
  \bibinfo{author}{\bibfnamefont{O.}~\bibnamefont{Philipsen}},
  \bibinfo{journal}{Nucl. Phys.} \textbf{\bibinfo{volume}{B673}},
  \bibinfo{pages}{170} (\bibinfo{year}{2003}), \eprint{hep-lat/0307020}.

\bibitem[{\citenamefont{Allton et~al.}(2005)}]{Allton:2005gk}
\bibinfo{author}{\bibfnamefont{C.~R.} \bibnamefont{Allton}}
  \bibnamefont{et~al.}, \bibinfo{journal}{Phys. Rev.}
  \textbf{\bibinfo{volume}{D71}}, \bibinfo{pages}{054508}
  (\bibinfo{year}{2005}), \eprint{hep-lat/0501030}.

\bibitem[{\citenamefont{Gavai and Gupta}(2005)}]{Gavai:2004sd}
\bibinfo{author}{\bibfnamefont{R.~V.} \bibnamefont{Gavai}} \bibnamefont{and}
  \bibinfo{author}{\bibfnamefont{S.}~\bibnamefont{Gupta}},
  \bibinfo{journal}{Phys. Rev.} \textbf{\bibinfo{volume}{D71}},
  \bibinfo{pages}{114014} (\bibinfo{year}{2005}), \eprint{hep-lat/0412035}.

\bibitem[{\citenamefont{Philipsen}(2006)}]{Philipsen:2005mj}
\bibinfo{author}{\bibfnamefont{O.}~\bibnamefont{Philipsen}},
  \bibinfo{journal}{Proceedings of Science} \textbf{\bibinfo{volume}{LAT2005}},
  \bibinfo{pages}{016} (\bibinfo{year}{2006}), \eprint{hep-lat/0510077}.

\bibitem[{\citenamefont{RHIC-Workshop}(2006)}]{CpWorkshop:2006}
\bibinfo{author}{\bibnamefont{RHIC-Workshop}} (\bibinfo{year}{2006}),
  \eprint{\\ https://www.bnl.gov/riken/QCDRhic/}.

\bibitem[{\citenamefont{Stephans}(2006)}]{Stephans:2006tg}
\bibinfo{author}{\bibfnamefont{G.~S.~F.} \bibnamefont{Stephans}}
  (\bibinfo{year}{2006}), \eprint{nucl-ex/0607030}.

\bibitem[{\citenamefont{Gazdzicki and Gorenstein}(1999)}]{Gazdzicki:1998vd}
\bibinfo{author}{\bibfnamefont{M.}~\bibnamefont{Gazdzicki}} \bibnamefont{and}
  \bibinfo{author}{\bibfnamefont{M.~I.} \bibnamefont{Gorenstein}},
  \bibinfo{journal}{Acta Phys. Polon.} \textbf{\bibinfo{volume}{B30}},
  \bibinfo{pages}{2705} (\bibinfo{year}{1999}), \eprint{hep-ph/9803462}.

\bibitem[{\citenamefont{Gazdzicki}(2005)}]{Gazdzicki:2005gs}
\bibinfo{author}{\bibfnamefont{M.}~\bibnamefont{Gazdzicki}}
  (\bibinfo{year}{2005}), \eprint{nucl-ex/0512034}.

\bibitem[{\citenamefont{Cleymans et~al.}(2006)\citenamefont{Cleymans, Oeschler,
  Redlich, and Wheaton}}]{Cleymans:2005xv}
\bibinfo{author}{\bibfnamefont{J.}~\bibnamefont{Cleymans}},
  \bibinfo{author}{\bibfnamefont{H.}~\bibnamefont{Oeschler}},
  \bibinfo{author}{\bibfnamefont{K.}~\bibnamefont{Redlich}}, \bibnamefont{and}
  \bibinfo{author}{\bibfnamefont{S.}~\bibnamefont{Wheaton}},
  \bibinfo{journal}{Phys. Rev.} \textbf{\bibinfo{volume}{C73}},
  \bibinfo{pages}{034905} (\bibinfo{year}{2006}), \eprint{hep-ph/0511094}.

\bibitem[{\citenamefont{Costa}(2007)}]{Costa:2007nc}
\bibinfo{author}{\bibfnamefont{P.}~\bibnamefont{Costa}}, \bibinfo{journal}{AIP
  Conf. Proc.} \textbf{\bibinfo{volume}{892}}, \bibinfo{pages}{255}
  (\bibinfo{year}{2007}), \eprint{hep-ph/0702232}.

\bibitem[{\citenamefont{Csernai et~al.}(2006)\citenamefont{Csernai, Kapusta,
  and McLerran}}]{Csernai:2006zz}
\bibinfo{author}{\bibfnamefont{L.~P.} \bibnamefont{Csernai}},
  \bibinfo{author}{\bibfnamefont{J.~I.} \bibnamefont{Kapusta}},
  \bibnamefont{and} \bibinfo{author}{\bibfnamefont{L.~D.}
  \bibnamefont{McLerran}}, \bibinfo{journal}{Phys. Rev. Lett.}
  \textbf{\bibinfo{volume}{97}}, \bibinfo{pages}{152303}
  (\bibinfo{year}{2006}), \eprint{nucl-th/0604032}.

\bibitem[{\citenamefont{Lacey et~al.}(2007)}]{Lacey:2006bc}
\bibinfo{author}{\bibfnamefont{R.~A.} \bibnamefont{Lacey}}
  \bibnamefont{et~al.}, \bibinfo{journal}{Phys. Rev. Lett.}
  \textbf{\bibinfo{volume}{98}}, \bibinfo{pages}{092301}
  (\bibinfo{year}{2007}), \eprint{nucl-ex/0609025}.

\bibitem[{\citenamefont{Chen and Nakano}(2006)}]{Chen:2006ig}
\bibinfo{author}{\bibfnamefont{J.-W.} \bibnamefont{Chen}} \bibnamefont{and}
  \bibinfo{author}{\bibfnamefont{E.}~\bibnamefont{Nakano}}
  (\bibinfo{year}{2006}), \eprint{hep-ph/0604138}.

\bibitem[{\citenamefont{Kovtun et~al.}(2005)\citenamefont{Kovtun, Son, and
  Starinets}}]{Kovtun:2004de}
\bibinfo{author}{\bibfnamefont{P.}~\bibnamefont{Kovtun}},
  \bibinfo{author}{\bibfnamefont{D.~T.} \bibnamefont{Son}}, \bibnamefont{and}
  \bibinfo{author}{\bibfnamefont{A.~O.} \bibnamefont{Starinets}},
  \bibinfo{journal}{Phys. Rev. Lett.} \textbf{\bibinfo{volume}{94}},
  \bibinfo{pages}{111601} (\bibinfo{year}{2005}), \eprint{hep-th/0405231}.

\bibitem[{\citenamefont{Chen et~al.}(2007)\citenamefont{Chen, Huang, Li,
  Nakano, and Yang}}]{Chen:2007jq}
\bibinfo{author}{\bibfnamefont{J.-W.} \bibnamefont{Chen}},
  \bibinfo{author}{\bibfnamefont{M.}~\bibnamefont{Huang}},
  \bibinfo{author}{\bibfnamefont{Y.-H.} \bibnamefont{Li}},
  \bibinfo{author}{\bibfnamefont{E.}~\bibnamefont{Nakano}}, \bibnamefont{and}
  \bibinfo{author}{\bibfnamefont{D.-L.} \bibnamefont{Yang}}
  (\bibinfo{year}{2007}), \eprint{arXiv:0709.3434 [hep-ph]}.

\bibitem[{\citenamefont{Issah and Taranenko}(2006)}]{Issah:2006qn}
\bibinfo{author}{\bibfnamefont{M.}~\bibnamefont{Issah}} \bibnamefont{and}
  \bibinfo{author}{\bibfnamefont{A.}~\bibnamefont{Taranenko}}
  (\bibinfo{collaboration}{PHENIX}) (\bibinfo{year}{2006}),
  \eprint{nucl-ex/0604011}.

\bibitem[{\citenamefont{Adare}(2006)}]{Adare:2006ti}
\bibinfo{author}{\bibfnamefont{A.}~\bibnamefont{Adare}}
  (\bibinfo{collaboration}{PHENIX}) (\bibinfo{year}{2006}),
  \eprint{nucl-ex/0608033}.

\bibitem[{\citenamefont{Lacey and Taranenko}(2006)}]{Lacey:2006pn}
\bibinfo{author}{\bibfnamefont{R.~A.} \bibnamefont{Lacey}} \bibnamefont{and}
  \bibinfo{author}{\bibfnamefont{A.}~\bibnamefont{Taranenko}}
  (\bibinfo{year}{2006}), \eprint{nucl-ex/0610029}.

\bibitem[{eta()}]{etas_uncertainty}
\bibinfo{note}{This estimate is consistent with the observed $v_2$ for
  non-photonic electrons and the universal scaling of $v_2$ for D mesons
  \cite{Lacey:2006pn,Adare:2006nq}.}

\bibitem[{com()}]{compare_etas}
\bibinfo{note}{This value for $\frac{\eta}{s}$ is in good agreement with the
  experimentally based estimates of Teaney, Gavin and PHENIX
  \cite{Teaney:2003kp,Gavin:2006xd,Adare:2006nq} and the theoretical estimates
  of Gyulassy and Shuryak \cite{Hirano:2005wx,Gelman:2006xw}. All of these
  estimates contrast the predictions of pertubative QCD
  \cite{Teaney:2003kp,Arnold:2000dr}.}

\bibitem[{\citenamefont{Karsch et~al.}(2001)\citenamefont{Karsch, Laermann, and
  Peikert}}]{Karsch:2000kv}
\bibinfo{author}{\bibfnamefont{F.}~\bibnamefont{Karsch}},
  \bibinfo{author}{\bibfnamefont{E.}~\bibnamefont{Laermann}}, \bibnamefont{and}
  \bibinfo{author}{\bibfnamefont{A.}~\bibnamefont{Peikert}},
  \bibinfo{journal}{Nucl. Phys.} \textbf{\bibinfo{volume}{B605}},
  \bibinfo{pages}{579} (\bibinfo{year}{2001}), \eprint{hep-lat/0012023}.

\bibitem[{\citenamefont{Adler et~al.}(2005{\natexlab{a}})}]{Adler:2004cj}
\bibinfo{author}{\bibfnamefont{S.~S.} \bibnamefont{Adler}}
  \bibnamefont{et~al.}, \bibinfo{journal}{Phys. Rev. Lett.}
  \textbf{\bibinfo{volume}{94}}, \bibinfo{pages}{232302}
  (\bibinfo{year}{2005}{\natexlab{a}}).

\bibitem[{\citenamefont{Adler et~al.}(2005{\natexlab{b}})}]{Adler:2004zn}
\bibinfo{author}{\bibfnamefont{S.~S.} \bibnamefont{Adler}} \bibnamefont{et~al.}
  (\bibinfo{collaboration}{PHENIX}), \bibinfo{journal}{Phys. Rev.}
  \textbf{\bibinfo{volume}{C71}}, \bibinfo{pages}{034908}
  (\bibinfo{year}{2005}{\natexlab{b}}).

\bibitem[{\citenamefont{Margetis et~al.}(1995)}]{Margetis:1994tt}
\bibinfo{author}{\bibfnamefont{S.}~\bibnamefont{Margetis}} \bibnamefont{et~al.}
  (\bibinfo{collaboration}{NA49}), \bibinfo{journal}{Phys. Rev. Lett.}
  \textbf{\bibinfo{volume}{75}}, \bibinfo{pages}{3814} (\bibinfo{year}{1995}).

\bibitem[{\citenamefont{Hirano et~al.}(2006)}]{Hirano:2005xf}
\bibinfo{author}{\bibfnamefont{T.}~\bibnamefont{Hirano}} \bibnamefont{et~al.},
  \bibinfo{journal}{Phys. Lett.} \textbf{\bibinfo{volume}{B636}},
  \bibinfo{pages}{299} (\bibinfo{year}{2006}).

\bibitem[{\citenamefont{Itakura et~al.}(2007)\citenamefont{Itakura, Morimatsu,
  and Otomo}}]{Itakura:2007mx}
\bibinfo{author}{\bibfnamefont{K.}~\bibnamefont{Itakura}},
  \bibinfo{author}{\bibfnamefont{O.}~\bibnamefont{Morimatsu}},
  \bibnamefont{and} \bibinfo{author}{\bibfnamefont{H.}~\bibnamefont{Otomo}}
  (\bibinfo{year}{2007}), \eprint{arXiv:0711.1034 [hep-ph]}.

\bibitem[{\citenamefont{Landau and Lifschitz}(1986)}]{Landdau1986}
\bibinfo{author}{\bibfnamefont{L.~D.} \bibnamefont{Landau}} \bibnamefont{and}
  \bibinfo{author}{\bibfnamefont{E.~M.} \bibnamefont{Lifschitz}},
  \emph{\bibinfo{title}{Fluid Mechanics}}
  (\bibinfo{publisher}{Butterworth-Heinemann}, \bibinfo{year}{1986}).

\bibitem[{\citenamefont{Eckart}(1940)}]{Eckart1940}
\bibinfo{author}{\bibfnamefont{C.}~\bibnamefont{Eckart}},
  \bibinfo{journal}{Phys. Rev.} \textbf{\bibinfo{volume}{58}},
  \bibinfo{pages}{919} (\bibinfo{year}{1940}).

\bibitem[{\citenamefont{Adare et~al.}(2006)}]{Adare:2006nq}
\bibinfo{author}{\bibfnamefont{A.}~\bibnamefont{Adare}} \bibnamefont{et~al.}
  (\bibinfo{year}{2006}), \eprint{nucl-ex/0611018}.

\bibitem[{\citenamefont{Teaney}(2003)}]{Teaney:2003kp}
\bibinfo{author}{\bibfnamefont{D.}~\bibnamefont{Teaney}},
  \bibinfo{journal}{Phys. Rev.} \textbf{\bibinfo{volume}{C68}},
  \bibinfo{pages}{034913} (\bibinfo{year}{2003}), \eprint{nucl-th/0301099}.

\bibitem[{\citenamefont{Gavin and Abdel-Aziz}(2006)}]{Gavin:2006xd}
\bibinfo{author}{\bibfnamefont{S.}~\bibnamefont{Gavin}} \bibnamefont{and}
  \bibinfo{author}{\bibfnamefont{M.}~\bibnamefont{Abdel-Aziz}}
  (\bibinfo{year}{2006}), \eprint{nucl-th/0606061}.

\bibitem[{\citenamefont{Hirano and Gyulassy}(2006)}]{Hirano:2005wx}
\bibinfo{author}{\bibfnamefont{T.}~\bibnamefont{Hirano}} \bibnamefont{and}
  \bibinfo{author}{\bibfnamefont{M.}~\bibnamefont{Gyulassy}},
  \bibinfo{journal}{Nucl. Phys.} \textbf{\bibinfo{volume}{A769}},
  \bibinfo{pages}{71} (\bibinfo{year}{2006}), \eprint{nucl-th/0506049}.

\bibitem[{\citenamefont{Gelman et~al.}(2006)\citenamefont{Gelman, Shuryak, and
  Zahed}}]{Gelman:2006xw}
\bibinfo{author}{\bibfnamefont{B.~A.} \bibnamefont{Gelman}},
  \bibinfo{author}{\bibfnamefont{E.~V.} \bibnamefont{Shuryak}},
  \bibnamefont{and} \bibinfo{author}{\bibfnamefont{I.}~\bibnamefont{Zahed}}
  (\bibinfo{year}{2006}), \eprint{nucl-th/0601029}.

\bibitem[{\citenamefont{Arnold et~al.}(2000)\citenamefont{Arnold, Moore, and
  Yaffe}}]{Arnold:2000dr}
\bibinfo{author}{\bibfnamefont{P.}~\bibnamefont{Arnold}},
  \bibinfo{author}{\bibfnamefont{G.~D.} \bibnamefont{Moore}}, \bibnamefont{and}
  \bibinfo{author}{\bibfnamefont{L.~G.} \bibnamefont{Yaffe}},
  \bibinfo{journal}{JHEP} \textbf{\bibinfo{volume}{11}}, \bibinfo{pages}{001}
  (\bibinfo{year}{2000}).

\end{thebibliography}
\end{document}